\documentclass[aps,reprint,prd,nofootinbib]{revtex4-1}

\usepackage{bm}
\usepackage{graphicx} %
\usepackage{amsmath} %
\usepackage{amssymb} %
\usepackage{url}
\usepackage{subfigure} %
\usepackage{float} %
\usepackage{upgreek} %
\usepackage{multirow} %
\usepackage{color} %
\usepackage{lineno} %
\usepackage{hyperref} %
\usepackage{booktabs}
\usepackage{placeins}
\usepackage[usenames,dvipsnames]{xcolor}
\hypersetup{colorlinks=true, urlcolor=black, linkcolor=blue, citecolor=red} %

\allowdisplaybreaks[4]
\usepackage[utf8]{inputenc}
\usepackage{autobreak}

\begin{document}

\title{Induced superhorizon tensor perturbations from anisotropic non-Gaussianity}

\author{Atsuhisa Ota}
\email{a.ota@damtp.cam.ac.uk}
\affiliation{Department of Applied Mathematics and Theoretical Physics, University of Cambridge, Center for Mathematical Science, Wilberforce Rd, Cambridge, CB3 0WA, UK}

\date{\today}

\begin{abstract}
We study cosmological tensor perturbations induced by second-order scalar perturbations in the presence of \textit{anisotropic} non-Gaussianity.
This class of induced tensor modes arises on superhorizon scales through the intrinsic quadrupole coupling between long modes and short modes.
Scalar perturbations on all scales from the inflationary Hubble radius to the Silk damping scale at recombination contribute to the induced tensor powerspectrum at the cosmic microwave background~(CMB) scale.
In addition, the induced tensor spectrum becomes almost scale-invariant.
The former property suggests that measurements of the CMB offer a test of tiny scale physics.
However, the latter implies the secondary effect may contaminate the primordial tensor spectrum, which tells us the energy scale of inflation.
We derive the induced tensor modes originated from two concrete examples of anisotropic non-Gaussianity; statistically anisotropic scalar non-Gaussianity and scalar-scalar-tensor non-Gaussianity, and discuss observational consequences of extremely short scale physics.
Also, we comment on various possibilities of enhancing the induced spectrum with nonstandard early Universe physics.

\end{abstract}

\maketitle


\section{Introduction}

A common feature of gravitational waves sourced \textit{after} inflation is infrared behavior of the powerspectrum~\cite{Baumann:2007zm,Caprini:2009fx,Cai:2019cdl}.
The induced spectrum is always suppressed on superhorizon scales because some physical process causally generates the gravitational waves.
For example, a quadratic source of scalar perturbations induces the dimensionless tensor powerspectrum, which scales as $q^3$ in the $q\to 0$ limit.
Consequently, even if catastrophic gravitational wave productions happened inside the tiny horizons in the early Universe, we would not see the remnants in the low frequency tale of the induced tensor powerspectrum that is related to $B$-mode polarization of the cosmic microwave background~(CMB).

In the above example, the previous works mostly assumed Gaussian scalar perturbations for the initial condition, which does not allow mode coupling of different Fourier modes.
Causal processes secondarily produce additional mode couplings between subhorizon modes, but superhorizon correlations never arise.
Thus, Gaussianity of the initial scalar perturbations is a reason for the scaling.
On the other hand, primordial non-Gaussianity, in particular, the local shape gives intrinsic mode mixing between long modes and short modes at second-order as the primordial bispectrum is nonzero in the squeezed limit.
This property means that subhorizon modes couple to the superhorizon modes at second-order.
Therefore, including this type of initial condition, we will see entirely different infrared scaling of the induced tensor powerspectrum.
It would be interesting if the induced superhorizon modes carry cosmological information at tiny scales because we can test it, using the recent polarization B-mode measurements~(see Refs.~\cite{Sayre:2019dic, Ade:2018iql, Adachi:2019mjv, Akrami:2018odb} for recent efforts of B-mode measurements).
Refs.~\cite{Nakama:2016gzw, Cai:2018dig, Unal:2018yaa, Unal:2018yaa} studied primordial non-Gaussianity for the induced gravitational waves, but no one has considered the generation of superhorizon tensor perturbations.
Indeed, standard isotropic non-Gaussianity never induces the superhorizon tensor modes, and we will show quadrupole anisotropy in non-Gaussianity is essential.

In this paper, we compute induced tensor perturbations for statistically anisotropic scalar non-Gaussianity~\cite{Bartolo:2012sd,Shiraishi:2013vja,Abolhasani:2013zya,Shiraishi:2013oqa,Arkani-Hamed:2015bza,Lee:2016vti,Shiraishi:2016mok,Baumann:2017jvh,Bordin:2018pca} and scalar-scalar-tensor non-Gaussianity~\cite{Maldacena:2002vr,Seery:2008ax,Shiraishi:2010kd,Gao:2012ib,Franciolini:2017ktv,Domenech:2017kno,Biagetti:2017viz}.
The former correlation functions yield in inflationary models in the presence of the spinning fields, which break background isotropies in the early Universe~\cite{Arkani-Hamed:2015bza, Lee:2016vti, Shiraishi:2016mok, Baumann:2017jvh, Bordin:2018pca}.
The latter case is more common even for single-field inflation models~\cite{Maldacena:2002vr,Seery:2008ax,Shiraishi:2010kd,Gao:2012ib}, and one expects enhancements in some models with massive spin-2 fields~\cite{Domenech:2017kno,Biagetti:2017viz}.
We discuss the observational consequences of these non-Gaussian initial conditions with various nonstandard early Universe physics such as a specific reheating scenario and primordial blackhole formation, which enhance the secondary tensor perturbations.

\medskip

\medskip
This paper is organized as follows.
First, we give a brief introduction to mode coupling between long modes and short modes in the presence of local non-Gaussianity in section~\ref{ismdtpng}.
We introduce operator product expansion~(OPE) of cosmological perturbations, which is a useful mathematical tool for the soft limit calculations.
Then, we review an example of the mode coupling effects in the case of scalar perturbations.
In section~\ref{nonlinearpert}, we present a second-order cosmological perturbation theory to compute induced tensor perturbations.
We derive a superhorizon solution for non-Gaussian initial conditions and study the evolution of the source functions in section~\ref{sec:window}.
We show that the superhorizon tensor perturbations arise if the OPE coefficients have quadrupole anisotropy.
In section~\ref{initial:cd}, we give two examples of the anisotropic OPE coefficients: statistically anisotropic non-Gaussianity and scalar-scalar-tensor non-Gaussianity.
We discuss the observational consequences of these two examples in sections~\ref{tsr:sa} and \ref{tsr:sst}.
Then we give a summary and conclusions in the final section.

\section{Mode coupling effects and primordial non-Gaussianity}\label{ismdtpng}

In this section, we explain the role of primordial non-Gaussianity for the mode coupling effects of cosmological perturbations.
We will also give an example of observable mode coupling effects in the case of scalar perturbations.

\subsection{Operator product expansion for cosmological perturbations}

In the standard framework of cosmological perturbation theory, the inflationary Universe sets initial conditions of cosmological perturbations on superhorizon scales.
These modes are constant on superhorizon scales, re-enter the horizon as the Universe expands, and evolve to form various structures in the present Universe~(e.g.~Refs~\cite{Mukhanov:2005sc, Dodelson:2003ft} for reviews).
While they are initially almost linear Gaussian perturbations, causal processes secondarily produce nonlinearity on subhorizon scales.
Even in the radiation era where the density perturbations do not grow, there exist nonlinear effects.
For instance, we can describe friction heat due to shear in a viscous photon-baryon plasma in second-order cosmological perturbation theory.
We see the linear CMB temperature anisotropy is damping at multipole $\ell>\mathcal O(100)$ in the observed angular powerspectrum.
We only observe this damping feature at first order, but the friction heat arises and deforms the blackbody spectrum at second-order~\cite{Hu:1994bz, Pitrou:2009bc, Chluba:2011hw, Chluba:2012gq, Ota:2016esq}.  
Not only the scalar perturbations but also the vector and tensor perturbations arise at second-order. 
The induced second-order vorticity in the photon-baryon plasma during recombination leads to magnetic fields~\cite{Matarrese:2004kq,Takahashi:2005nd,Matarrese:2004kq,Kobayashi:2007wd,Fenu:2010kh,Nalson:2013jya,Saga:2015bna,Fidler:2015kkt,Carrilho:2019qlb}, and the quadrupole anisotropy at second-order scalar perturbations produce the tensor perturbations~\cite{Matarrese:1993zf,Matarrese:1997ay,Mollerach:2003nq,Ananda:2006af,Baumann:2007zm,Saito:2008jc,Saito:2009jt,Assadullahi:2009nf,Caprini:2009fx,Alabidi:2013wtp,Nakama:2016gzw,Garcia-Bellido:2017aan,Kohri:2018awv,Unal:2018yaa,Unal:2018yaa,Inomata:2018epa,Cai:2019amo,Inomata:2019zqy,Inomata:2019ivs,Inomata:2018epa,Cai:2019elf,Cai:2019cdl,Gong:2019mui,Domenech:2019quo,Yuan:2019fwv,DeLuca:2019ufz,Inomata:2019yww}.
A common thing among these secondary effects is that physical processes induce these effects.
Hence the powerspectra of these secondary effects are significant, in principle, at causal scales.
The dimensionless powerspectra of them scale as $q^{\gamma}$ with $\gamma>0$ on superhorizon scales~\cite{Baumann:2007zm,Caprini:2009fx,Cai:2019cdl,Pajer:2012vz,Saga:2015bna,Fidler:2015kkt}, in the variety of examples.
Thus the secondary cosmological perturbations are not produced on superhorizon scales $q\to 0$, which makes sense in terms of causality.

\medskip
So far we mentioned secondary effects originated from a Gaussian primordial perturbation $\xi$, which we write the powerspectrum in the following way:
\begin{align}
    \langle \xi(\mathbf q)\xi(\mathbf q')\rangle = (2\pi)^3 \delta^{(3)}(\mathbf q+\mathbf q')P(q).\label{def:pow:xi}
\end{align}
$\xi$ can be, for example, the adiabatic or isocurvature perturbations.
Various inflationary models predict $q^3 P(q)$ is almost scale-invariant with a small tilt.
The delta function appears because of background homogeneity during inflation.
Thus, Gaussian perturbations only couple to the same Fourier modes.
On the other hand, many inflationary models also predict intrinsic nonlinearity in cosmological perturbations.
For example, we write local form non-Gaussianity as
\begin{align}
        \langle \xi(\mathbf q)\xi(\mathbf k_1)\xi(\mathbf k_2) \rangle     
        = (2\pi)^3\delta(\mathbf q+\mathbf k_1+\mathbf k_2) B_\xi(q,k_1,k_2),
    \label{def:bs}                
\end{align}
where we have defined 
\begin{align}
   B_\xi(q,k_1,k_2)\equiv f_{\rm NL} \left(P(q)P(k_1)+2~{\rm perms.}\right).\label{def:bp:2}
\end{align}
In contrast to Eqs.~\eqref{def:pow:xi}, \eqref{def:bs} tells us $\xi$ couples to the various Fourier modes at second-order.
Moreover, Fourier modes are correlated even for a squeezed limit where one of the three modes is on superhorizon scales.
This kind of correlation is possible because some nonlinear interaction induces them during inflation~\cite{Maldacena:2002vr}.
Thus, thanks to local form primordial non-Gaussianity, the superhorizon perturbations couple to the subhorizon ones a priori in the Universe after inflation. 

\medskip
Operator product expansion~(OPE) of cosmological perturbations gives us a more intuitive picture of this sort of mode coupling.
Eqs.~\eqref{def:pow:xi} to \eqref{def:bp:2} suggest that we may expand a momentum space operator product in the following way:
\begin{align}
\begin{split}
    &\xi\left( \mathbf k+\frac{\mathbf q}{2}\right)\xi\left(-\mathbf k+\frac{\mathbf q}{2}\right) \\
    = &P(k)(2\pi)^3\delta^{(3)}(\mathbf q)+ C_\xi(\mathbf k,\mathbf q) \xi(\mathbf q)+\cdots+\mathcal O(q/k),
\end{split}    
\label{OPE}
    \end{align}
where $\cdots$ imply some possible operators for given initial conditions.
Taking the expectation value of both sides of Eq.\eqref{OPE}, and taking $q/k\to 0$ limit, we immediately find the first term is consistent with Eq.~\eqref{def:pow:xi}.
Multiplying $\xi(\mathbf q')$ on the both sides, one finds the matching condition for $C_{\xi}$:
\begin{align}
\begin{split}
    &\left\langle \xi(\mathbf q') \xi\left( \mathbf k+\frac{\mathbf q}{2}\right)\xi\left(-\mathbf k+\frac{\mathbf q}{2}\right) \right\rangle  \\
    = &(2\pi)^3\delta^{(3)}(\mathbf q+\mathbf q')C_\xi(\mathbf k,\mathbf q) P(q)+\cdots+\mathcal O(q/k).
\end{split}
\end{align}
The left hand side~(LHS) is given in Eqs.~\eqref{def:bs} and \eqref{def:bp:2}, so we find 
\begin{align}
    C_\xi = 2f_{\rm NL} P(k).\label{matching:1}
\end{align}
Thus, non-vanishing $C_\xi$ means mode coupling between short modes and long modes.
Depending on the types of primordial non-Gaussianity, we have a variety of possibilities of OPE.
We will discuss concrete examples of them in section~\ref{initial:cd}.

\subsection{An example of the mode coupling effect: CMB spectral distortion anisotropy}\label{cmb:mudistortion}
Next, we review an example of the actual cosmological observables related to the above mode coupling effect: CMB spectral distortion anisotropy.
The energy spectrum of the CMB is a perfect blackbody with small anisotropies because the early Universe was dense radiation fluid, and photon interactions rapidly realize local chemical equilibrium states.
However, precise numerical simulations revealed that the early Universe was out of chemical equilibrium after the redshift $z\simeq 2\times 10^6$ when the double Compton scattering becomes inefficient~\cite{1982A&A...107...39D,1991A&A...246...49B}.
After that, the Compton scattering dominates photon interactions, which conserves the number of photons; therefore, the thermalization of nontrivial energy injection to the local blackbody will not only change the local temperature but also induce the nonzero chemical potential.
This process continues until $z\simeq 5\times 10^4$ when the Universe departs from kinetic equilibrium.
A possible source of this additional energy injection is friction heat that arises because of the acoustic oscillation of a viscous photon-baryon plasma~\cite{Hu:1994bz, Chluba:2011hw, Chluba:2012gq}.
The friction heat is second-order in the cosmological perturbations, and hence the induced chemical potential $\mu$ can be generally written as~\cite{Pajer:2013oca} 
\begin{align}
\begin{split}
    \mu(\eta, \mathbf q;\hat p) =    &\int \frac{d^3k_1d^3k_2}{(2\pi)^6}(2\pi)^3\delta^{(3)}(\mathbf k_1+\mathbf k_2-\mathbf q) \\
    &\times \tilde \mu (\eta,\mathbf k_1,\mathbf k_2;\hat p)\xi(\mathbf k_1)\xi(\mathbf k_2),
\end{split}
    \label{def:mu:NL}
\end{align}
where $\tilde \mu (\eta,\mathbf k_1,\mathbf k_2;\hat p)$ is a transfer function in second-order perturbation theory that we find by solving fluid dynamics with gravity.
Note that $\hat p$ is the direction of photons.
Introducing a new coordinate
\begin{align}
    \mathbf k_1\equiv \mathbf k+\frac{\mathbf q'}{2},~~\mathbf k_2\equiv -\mathbf k+\frac{\mathbf q'}{2},\label{sq:coordinate:1}
\end{align}
we integrate Eq.~\eqref{def:mu:NL} with respect to $\mathbf q'$, and one finds
\begin{align}
\begin{split}
    \mu(\eta,\mathbf q;\hat p)     =&\int \frac{d^3k}{(2\pi)^3}\tilde \mu \left(\eta, \mathbf k+\frac{\mathbf q}{2},-\mathbf k+\frac{\mathbf q}{2};\hat p\right)   \\
    & \times  \xi\left( \mathbf k+\frac{\mathbf q}{2}\right)\xi\left(-\mathbf k+\frac{\mathbf q}{2}\right).
\end{split}
    \label{mu:skemat}
\end{align}
Employing Eq.~\eqref{def:pow:xi}, we find an averaged $\mu$ parameter
\begin{align}
    \langle \mu(\eta,\mathbf x;\hat p) \rangle =  \int \frac{d^3k}{(2\pi)^3}\tilde \mu \left(\eta, \mathbf k,-\mathbf k;\hat p\right)P(k),\label{def:monopolemu}
\end{align}
where real space position $\mathbf x$ can be arbitrary.
On the other hand, with Eqs.~\eqref{OPE} and \eqref{matching:1}, the fluctuation of the chemical potential $\delta \mu \equiv \mu - \langle \mu\rangle$ in Fourier space becomes~\cite{Pajer:2012vz,Emami:2015xqa}
\begin{align}
    \delta \mu(\eta,\mathbf q;\hat p)     =& 2f_{\rm NL}\langle \mu \rangle \xi(\mathbf q) +\mathcal O(q/k).\label{mu:skemat:f}
\end{align}
Let $k_0$ be the horizon scale. 
Then the transfer function $\tilde \mu(\eta, \mathbf k_1,\mathbf k_2;\hat p)$ will mask $k_1,k_2<k_0$ modes because physical dissipative process happens only on subhorizon scales.
Hence, for the $q\ll k_0$ modes, we safely truncate the gradient corrections in Eq.~\eqref{mu:skemat:f}.
Thus, short-wavelength perturbations produce $\mu$ fluctuations on top of the long-wavelength linear perturbations through primordial non-Gaussianity.
The monopole spectral distortion anisotropy is constant before it enters the horizon as usual density perturbations.
This process does not violate causality because nonlinear interactions during inflation induce the mode coupling between long modes and short modes.

\begin{figure}
\centering
  \includegraphics[width=0.8\linewidth]{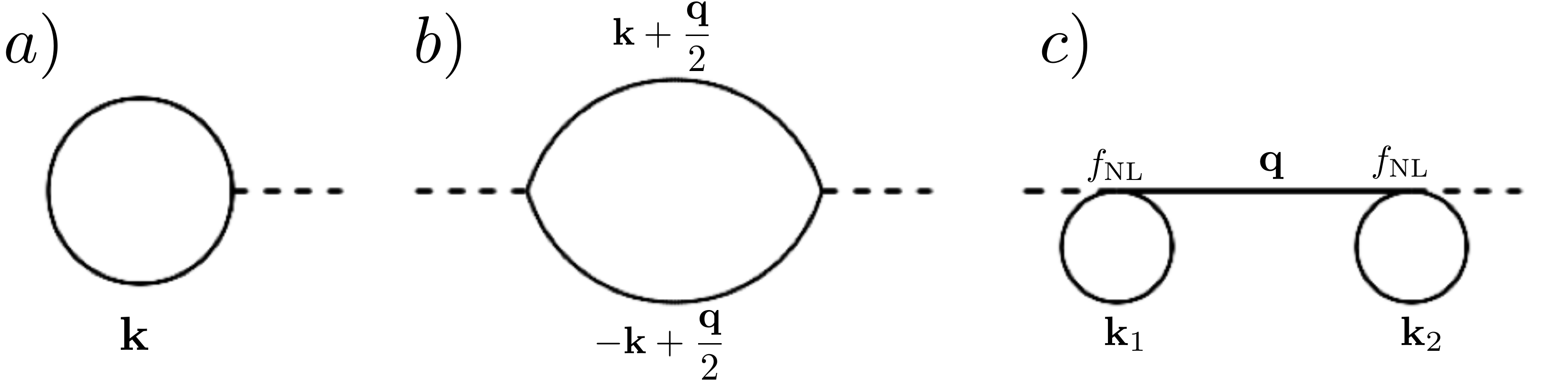}
  \caption{
Diagrams of the secondary effects.
  $\tilde \mu$ and $P$ correspond to a dashed line and a solid line~(a propagator).
  We use $k$ and $q$ for a loop and an external momenta, respectively.
  $a$) $\langle \mu \rangle$ corresponds to a tadpole diagram but is not divergent because of $\tilde \mu$ in Eq.~\eqref{def:monopolemu}.
  $b$) $\langle \delta \mu_g\delta \mu_g\rangle$ is understood as a 1-loop diagram. 
  This diagram implies that a Gaussian induced $\mu\mu$ propagator $P^{\rm G}_{\mu\mu}$ is $q$ independent in $q/k\to 0$ limit. Hence $q^3 P^{\rm G}_{\mu\mu}\propto q^3$.
  Therefore superhorizon correlations are prohibited.
  $c$) $\langle \delta \mu_{ng}\delta \mu_{ng}\rangle$ corresponds to a 2-loop diagram. We can take the loop momenta $\mathbf k_1$ and $\mathbf k_2$ independently from the external momentum $q$. Therefore, we get a non-vanishing contribution for $q\to 0$.  
  }
  \label{loop}
\end{figure}

While the $\mu$ anisotropy in the CMB sky is tiny as its monopole is $\langle \mu \rangle \sim q^3 P\sim 10^{-9}$, the cross-correlation of $\delta \mu$ and the temperature perturbation~$\Theta$ is sensitive to $f_{\rm NL}$. 
This is because $\Theta\propto \xi$ so that $\langle \Theta \delta \mu \rangle \propto f_{\rm NL}P^2$~\cite{Pajer:2012vz}. 
Also, the $\mu$ auto powerspectrum is related to the collapsed limit trispectrum conventionally parameterized by $\tau_{\rm NL}$, which is related to the squeezed bispectrum as $\tau_{\rm NL}\geq f_{\rm NL}^2$~\cite{Suyama:2007bg}.
This inequality follows from all possible terms in Eq.~\eqref{OPE}, and hence $\tau_{\rm NL}=f^2_{\rm NL}$ for single field cases. 
One may wonder that the $\mu\mu$ auto powerspectrum is also sourced by the Gaussian perturbations, and the Gaussian contribution should be more significant than the non-Gaussian contribution because $\langle \delta \mu_g\delta \mu_g\rangle \sim P^2$ and $\langle \delta \mu_{ng}\delta \mu_{ng}\rangle \sim \tau_{\rm NL}P^3$.
As we show in Fig.~\ref{loop}, we may consider these $\mu\mu$ correlation functions are 1-loop and 2-loop diagrams, respectively.
Therefore, naively, we expect the 1-loop diagram is more significant than the 2-loop contribution~(see caption of Fig.~\ref{loop} for details of diagrammatic explanations).

Indeed, $\mathcal O(q/k)$ corrections contain the Gaussian contribution, and hence the 1-loop contribution is suppressed for $q/k\to 0$.
In Ref.~\cite{Pajer:2012vz}, the angular powerspectrum of $\mu$ is evaluated for both Gaussian and non-Gaussian initial conditions as 
\begin{align}
    \ell^2C^{\mu\mu,g}_\ell &\sim \ell^2 10^{-29},\\
    \ell^2 C^{\mu\mu,ng}_\ell& \sim 10^{-23}\tau_{\rm NL}.
\end{align}
The non-Gaussian contribution is dominant even for $\mathcal O(1)$ of $\tau_{\rm NL}$ on the CMB scale $\mathcal O(10^{0})<\ell_{\rm obs.}<\mathcal O(10^{3})$.
Thus, we can measure small scale non-Gaussianity by observing the large scale $\mu$ anisotropies~\cite{Pajer:2012vz,Ganc:2012ae,Pajer:2013oca,Biagetti:2013sr,Ota:2014iva,Emami:2015xqa,Shiraishi:2015lma,Bartolo:2015fqz,Bartolo:2015fqz,Shiraishi:2016hjd,Ota:2016mqd,Ravenni:2017lgw,Chluba:2016aln,Khatri:2015tla,Cabass:2018jgj}.

Although the $\mu$ anisotropies are damping as the CMB temperature anisotropies on small scales, let us ignore this effect and extrapolate the angular powerspectrum to $\ell_{\rm phys.}=10^{8}$, which corresponds to the physical scales of $\mu$ generation at the redshift $z\sim 10^6$:
\begin{align}
     \ell^2C^{\mu\mu,g}_\ell \Big|_{\ell=\ell_{\rm phys.}} &\sim 10^{-13},\\
     \ell^2C^{\mu\mu,ng}_\ell \Big|_{\ell=\ell_{\rm phys.}} & \sim 10^{-23}\tau_{\rm NL}.
\end{align}
Thus, on the scale where $\mu$ is generated, the Gaussian contributions are dominant as expected.

\medskip
So far, we reviewed a concrete example of spectral distortion anisotropy, but we may generalize a conclusion of this section as follows: \textit{secondary quantities sourced by high $k$ modes will acquire long-wavelength correlations for local non-Gaussianity, and hence we may observe extremely short wavelength perturbations by observing low $k$ spectrum.}
Now, we notice that the induced tensor perturbations are similar to the spectral distortion anisotropy: quadratic sources on small scales generate them.
Then, can the induced tensor perturbations be produced on superhorizon scales due to primordial non-Gaussianity?

\section{Cosmological perturbation theory at second-order}\label{nonlinearpert}

As we described in the previous section, this paper aims to calculate the induced tensor perturbations in the presence of primordial non-Gaussianity.
More concretely, we are going to generalize Eq.~\eqref{mu:skemat:f} for the tensor perturbations and evaluate their powerspectrum.
Hence we perturb both gravity and matter up to second-order in cosmological perturbation theory.
In this section, we give a self-contained introduction to second-order cosmological perturbation theory and find the superhorizon solution of the induced tensor modes.
Then we introduce the induced tensor powerspectrum, which we will compare with observations.

\subsection{Gravity}\label{section:metric}

First of all, we write the metric tensor $g_{\mu\nu}$ in the following way:

\begin{align}
g_{00}=&-a^2e^{2{A}},\label{g00}\\
g_{0i}=&0,\\
g_{ij}=&a^2e^{2{D}}\delta_{ij}+a^2H_{ij},\label{gij}
\end{align}
where $a$ is the scale factor, and we have chosen conformal Newtonian gauge for the scalar perturbations.
The nonlinear metric perturbations can be expanded into $X \equiv \sum_{n=1}
X^{(n)}$ for $X=A$ and $D$ with $n$ being the order in primordial perturbations.  
$H_{ij}$ is traceless transverse secondary tensor perturbation or first-order primordial one. 
We ignore vector perturbations for simplicity.
Note that Refs.~\cite{Matarrese:1997ay, Hwang:2017oxa} discussed gauge dependence of the induced tensor perturbations, and Ref.~\cite{Tomikawa:2019tvi} recently evaluated them in various gauges and show that the induced effect increases in some gauges.
Thus, the second-order tensor perturbations depend on gauge conditions for the scalar perturbations.
In this paper, we work on the conformal Newtonian gauge, where one can identify the nonlinear instability with that in Newtonian gravity on subhorizon scales.

\medskip
Starting from Eqs.~\eqref{g00} to \eqref{gij}, we obtain each component of the Christoffel symbol
\begin{align}
\Gamma^\mu{}_{\nu\rho}\equiv \frac12g^{\mu\alpha}\left(\partial_\rho g_{\alpha\nu}+\partial_\nu g_{\alpha\rho}-\partial_\alpha g_{\nu\rho} \right)
\end{align}
as
\begin{align}
\Gamma^0{}_{00}=&\mathcal H+{A}',\\
\Gamma^0{}_{0i}=&\partial_i{A},\\
\Gamma^0{}_{ij}=&
e^{-2{A}+2{D}}\delta_{ij}\left[
\mathcal H+{D}'\right]+\mathcal H H_{ij}+\frac12H'_{ij},\\
\Gamma^{i}{}_{00}=&
e^{-2{D}+2{A}}\partial_i{A},\\
\Gamma^i{}_{0j}=&(\mathcal H+{D}')\delta_{ij} +\frac12 H'_{ij} ,\\
\begin{split}
    \Gamma^i{}_{jk}=&-\partial_i{D}\delta_{jk}+\partial_k{D}\delta_{ij}+\partial_j{D}\delta_{ik} \\
    &
-\frac12\partial_i H_{jk}+\frac12\partial_k H_{ij}+\frac12\partial_j H_{ik}.
\end{split}
\end{align}
A prime is a derivative with respect to the conformal time, and $\mathcal H\equiv a'/a$ is the comoving Hubble parameter.
Then the Ricci tensor 
\begin{align}
    R_{\mu\nu} = \partial_\alpha \Gamma^{\alpha}{}_{\mu\nu} - \partial_\mu \Gamma^{\alpha}{}_{\alpha\nu} + \Gamma^{\beta}{}_{\alpha\beta}\Gamma^{\alpha}{}_{\mu\nu} - \Gamma^{\alpha}{}_{\beta\mu}\Gamma^{\beta}{}_{\alpha\nu},
\end{align}
is obtained as
\begin{align}
    \begin{split}
        R_{00}=& -3\mathcal H'-3\mathcal H   {D}' +3\mathcal H A ' -3  {D}''+3 A '  {D}' \\ &-3  {D'}^2
     +e^{2 A -2  {D}}\left[\partial^2  A  +\partial  A \partial  {D} +(\partial  A )^2\right],    
    \end{split}
    \\
     R_{0i}=&-2\partial_i   {D}'+2\mathcal H\partial_i  A  +2  {D}'\partial_i A ,\\
     \begin{split}
              R_{ij}=&e^{2  {D}-2 A }\delta_{ij}\left[\mathcal H'+2 \mathcal H^2+  {D}''-\mathcal H A '\right. \\
     &\left.+5\mathcal H  {D}'-  {D}' A '+3  {D}'{}^2 \right]  \\
     & + \delta_{ij}\left[-\partial^2  {D}-\partial  {D}\partial  A - (\partial  {D} )^2 \right] \\
     & - \partial_i\partial_j  {D}  -   \partial_i\partial_j  A  +  \partial_i  {D} \partial_j  {D}  \\
     & - \partial_i A  \partial_j A   + \partial_i  {D} \partial_j A  + \partial_i A  \partial_j  {D}\\
     &+\frac12\left[H''_{ij}+2\mathcal H H'_{ij}-\nabla^2 H_{ij}\right].
     \end{split}
\end{align}
Finally, the Einstein tensor 
\begin{align}
    G^\mu{}_{\nu} = R^\mu{}_{\nu} -\frac{1}{2}\delta^\mu{}_{\nu} R^\alpha{}_{\alpha},
\end{align}
is given as 
\begin{align}
    \begin{split}
     G^0{}_{0}=&a^{-2}e^{-2 A }\left(-3\mathcal H^2-6\mathcal H  {D}'-3  {D}'{}^2\right) \\
     &+a^{-2}e^{-2  {D}}\left(2\partial^2   {D} + (\partial   {D})^2 \right) ,     
    \end{split}\\
     G^0{}_{i}=&a^{-2}e^{-2 A }  \left[-2\partial_i   {D}'+2\mathcal H\partial_i  A  +2  {D}'\partial_i A  \right] ,\label{G0i}\\
     \begin{split}
         G^i{}_{j}=& a^{-2} e^{-2 A }\delta_{ij}\left[-2\mathcal H'- \mathcal H^2-2  {D}''\right.
     \\
     &\left.+2\mathcal H A '-4\mathcal H  {D}'+2  {D}' A '-3  {D}'{}^2 \right]  \\
     & + a^{-2} e^{-2  {D}} \delta_{ij}\left[\partial^2 A  +\partial^2  {D} +  (\partial A  )^2 \right] \\
     & + a^{-2} e^{-2  {D}}\left[-\partial_i\partial_j  {D}  -   \partial_i\partial_j  A  +  \partial_i  {D} \partial_j  {D}  \right. \\
     &\left.- \partial_i A  \partial_j A   + \partial_i  {D} \partial_j A  + \partial_i A  \partial_j  {D}\right]\\
     & + \frac1{2a^2}\left[H''_{ij}+2\mathcal H H'_{ij}-\nabla^2 H_{ij}\right].\label{Gij}
     \end{split}
\end{align}
One finds a more detailed derivation, for example, in Ref.~\cite{Bartolo:2006cu}.

\subsection{Matter}
Next, we introduce the energy-momentum tensor.
Assuming the matter sector is a perfect fluid, we have the following energy-momentum tensor:
\begin{align}
    T_{\mu\nu} = (\rho+p)u_\mu u_\nu +pg_{\mu\nu},
\end{align}
where $u^\mu$ is time-like 4-velocity of the fluid, which satisfies $g_{\mu\nu}u^\mu u^\nu =-1$.
A static observer on the fluid measures the energy density $\rho$ and the pressure $p$.
Introducing the equation of state $p=w\rho$, we get
\begin{align}
    T^0{}_i =& \rho (1+w)u^0u_i,\label{T0i}\\
    T^i{}_j =& \rho (1+w)u^iu_j+ \rho w \delta_{ij}.\label{Tij}
\end{align}
Note that anisotropic stress is non negligible on small scales, but we drop it for simplicity.

\subsection{The Einstein equation}

Using the Einstein equation
\begin{align}
    G^\mu{}_{\nu} = \frac{1}{M_{\rm pl}^2} T^\mu{}_{\nu},~ M_{\rm pl}\equiv \frac{1}{\sqrt{8\pi G}},\label{Einstein:eq}
\end{align}    
we will compute the evolution of the tensor perturbation at second-order.
First of all, we find the evolution of the fluid velocity using Eqs.~\eqref{G0i}, \eqref{T0i} and \eqref{Einstein:eq} as
\begin{align}
    u_i=&\frac{2M_{\rm pl}^2}{u^0a^2\rho (1+w)}   \left[-\partial_i   {D}'+\mathcal H\partial_i  A   \right].
\end{align}
Then, we can write the spatial component of the energy momentum tensor \eqref{Tij}, using the metric perturbations, and one finds
\begin{align}
\begin{split}
    \frac{2 a^2}{M_{\rm pl}^2} T^i{}_j =&  \frac{8}{3}\frac{1}{\mathcal H^2 (1+w)} \\ &\times \partial_i \left[\mathcal H A-   {D}'   \right]\partial_j \left[\mathcal H A-   {D}'   \right] +\rho w \delta_{ij}.
\end{split}
     \label{EMT:source}
\end{align}
We also used Friedmann equation $3\mathcal H^2 M_{\rm pl}^2=a^2 \rho$, which follows from the background 00 component of the Einstein equation~\eqref{Einstein:eq}.
Combining this expression with Eq~\eqref{Gij}, we get
\begin{align}
     H''_{ij}+2\mathcal H H'_{ij}-\nabla^2 H_{ij}=S_{ij},\label{tensor:eq}
\end{align}
where the source term is given as
\begin{align}
    \begin{split}
    S_{ij}\equiv 
    &\frac{8}{3}\frac{1}{\mathcal H^2  (1+w)}   \partial_i \left[\mathcal H A-   {D}'   \right]\partial_j \left[\mathcal H A-   {D}'   \right] \\ 
&  +  2\left[-2  {D}\partial_i\partial_j  {D}  -2  {D}   \partial_i\partial_j  A  -  \partial_i  {D} \partial_j  {D} \right.\\
    &\left. + \partial_i A  \partial_j A  
     - \partial_i  {D} \partial_j A  - \partial_i A  \partial_j  {D}\right] + \cdots.
    \end{split}
\label{def:source}
    \end{align}
The dots imply the terms proportional to $\delta_{ij}$ or total derivatives, which do not contribute to the second-order tensor perturbations.

\medskip
We solve Eq.~\eqref{tensor:eq} in Fourier space where we expand cosmological perturbations into helicity.
In this paper, we define the Fourier transform of $X(\eta,\mathbf x)$ as follows:
\begin{align}
    X(\eta,\mathbf q) \equiv \int d^3x e^{-i\mathbf q\cdot \mathbf x}X(\eta,\mathbf x).\label{def:fourier}
\end{align}
For the adiabatic initial condition, we usually normalize the scalar perturbations by the curvature perturbation on the uniform density slice $\zeta$ as
\begin{align}
    A(\eta,\mathbf k) = \tilde A(\eta,k)\zeta(\mathbf k),~D(\eta,\mathbf k) = \tilde D(\eta,k)\zeta(\mathbf k),\label{def:transf}
\end{align}
where a tilde means a transfer function.
This $\zeta$ is a gauge-invariant and superhorizon conserved quantity; therefore, we commonly use it to relate some predictions of the early Universe models with physics in the late Universe.
We will discuss concrete expressions of the transfer functions in section~\ref{anal;transfsec}.

Tensor perturbations are expanded into
\begin{align}
H_{ij}(\eta,\mathbf q)=\sum_{s=\pm 2} H^{(s)}(\eta,\mathbf q)e^{(s)}_{ij}(\mathbf q),
\end{align}
where $\pm2$ represent the helicity $\pm 2$, and $e^{(s)}_{ij}$ is the gravitational wave polarization tensor whose normalization conditions are given as $e^{(s)}_{ij}e^{(s')*}_{ji}=\delta_{ss'}$.
More explicitly, setting the Fourier momentum $\mathbf q$ parallel to $z$-axis, we may choose a frame where 
\begin{align}
        e^{(\pm2)}_{ij}(\mathbf q) = \frac{1}{2}\begin{pmatrix}
  1& \pm i & 0\\
  \pm i & -1 & 0\\
  0 & 0 & 0
\end{pmatrix}.
\end{align}
Using the above helicity decomposition, we find the evolution equation of the tensor modes.
In Fourier space, multiplying $e^{(s)*}_{ij}$ and Eq.~\eqref{tensor:eq}, we obtain
\begin{align}
{H^{(s)}}''+2\mathcal H {H^{(s)}}'+q^2 {H^{(s)}} = e^{(s)*}_{ij}S_{ij}.\label{eom:tens:source}
\end{align}
The right hand side~(RHS) is written as a convolution of the linear perturbations as Eq.~\eqref{def:source} is a quadratic form of the metric perturbations.

\subsection{Superhorizon solution}\label{subsec:sup}

$H^{(s)}$ is composed of the intrinsic primordial tensor fluctuation produced during inflation $H^{(s)}_{\rm L}$ and the induced second-order perturbation $H^{(s)}_{\rm NL}$.
The former is a homogenous solution to Eq.~\eqref{eom:tens:source}, which we write
\begin{align}
    H^{(s)}_{\rm L}(\eta,\mathbf q) = \tilde H^{(s)}_{\rm L}(\eta, q)\xi^{(s)}(\mathbf q).
\end{align}
The transfer function $\tilde H^{(s)}_{\rm L}$ is set to unity on superhorizon scale $q\eta \to 0$.

We will find the latter solution by the Green's function method.
A Green's function of Eq.~\eqref{eom:tens:source} is easily found in $q\eta\to 0$ limit.
In this limit, we solve
\begin{align}
\begin{split}
    &G(\eta,\bar \eta)''+ 2\mathcal H(\eta) G(\eta,\bar \eta)' 
= \delta(\eta-\bar \eta).\label{def:Greensfunc}
\end{split}
\end{align}
It is straightforward to integrate this equation:
\begin{align}
    G(\eta,\bar \eta) = \int^\eta_{\bar \eta} d\eta' \frac{a(\bar \eta)^2}{a(\eta')^2}.\label{Green:func}
\end{align}
For an arbitrary constant $w~(\neq -1/3)$, the scale factor is given as $a\propto \eta^{\frac{2}{1+3w}}$.
The above Green's function depends on the equation of states at both $\bar \eta$ and $\eta'$, and $w$ is not necessarily the same for them.
For example, consider the induced tensor modes arise in the radiation era~($w(\bar \eta)=1/3$). 
Then we also need to follow their evolution in the late matter era~($w(\eta')=0$).  
Thus, we should split the $\eta'$ integral into several parts, depending on $w'\equiv w(\eta')$.
If $w'$ is constant for $\eta_n<\eta'<\eta_{n+1}$, one can expand Eq.~\eqref{Green:func} into
\begin{align}
    G(\eta,\bar \eta) = \sum_{n=1}^{N-1} G_n (\eta,\bar \eta),
\end{align}
where $\eta_1=\bar \eta$, $\eta_N=\eta$, and 
\begin{align}
    G_n (\eta,\bar \eta) &= \int^{\eta_{n+1}}_{\eta_n} d\eta' \Lambda_n^{-\epsilon'} \frac{\bar \eta^{\frac{4}{1+3\bar w}}}{{\eta'}^{\frac{4}{1+3w'}}},\\
    \epsilon' &\equiv \frac{12(w'-\bar w)}{(1+3\bar w)(1+3w')}.
\end{align}
Note that $\bar w=w(\bar \eta)$, and $\Lambda_n$ is a constant whose dimension is Mpc.
$\Lambda$ would be typically given by the conformal time at transition, so $\eta_{n-1}< \Lambda_n < \eta_{n}$.
Then we find
\begin{align}
     G_n (\eta,\bar \eta) =  \frac{1+3\bar w}{3(1-w')}\frac{\bar \eta }{\Lambda_n^{\epsilon'}}\left(\frac{\bar \eta^{\frac{3(1-\bar w)}{1+3\bar w}}}{\eta_n^{\frac{3(1-w')}{1+3w'}}}-\frac{\bar \eta^{\frac{3(1-\bar w)}{1+3\bar w}}}{\eta_{n+1}^{\frac{3(1-w')}{1+3w'}}}  \right).
\end{align}
Since we have $\bar \eta\le \eta_n<\eta_{n+1}$ and $w\le 1/3$, we obtain
\begin{align}
 G(\eta,\bar \eta) 
    = &\frac{1+3\bar w}{3(1-\bar w)} \bar \eta \left[1+ \mathcal O \left(\frac{\bar \eta^\gamma}{\Lambda_n^{\epsilon'} \eta_n^{\gamma-\epsilon'}} \right) \right],
\end{align}
where $\gamma>0$. 
This fact implies that the size of the leading term fixes when it arises at $\bar\eta$, and the induced superhorizon modes do not depend on $\eta$, i.e., the induced tensor perturbations are conserved on superhorizon scales.

The RHS of Eq.~\eqref{eom:tens:source} is also simplified for $q/k\to 0$.
Assuming the following OPE
\begin{align}
\begin{split}
    & \zeta\left( \mathbf k+\frac{\mathbf q}{2}\right)\zeta\left(-\mathbf k+\frac{\mathbf q}{2}\right)=P_\zeta(k)(2\pi)^3\delta^{(3)}(\mathbf q)\\  &+\sum_{s'=0,\pm2}C_{s'}(\mathbf k,\mathbf q) \xi^{(s')}(\mathbf q) +\mathcal O(q/k),    
\end{split}
\label{OPE2}
\end{align}
with $\xi^{(0)}\equiv \zeta$ and using Eqs.~\eqref{def:source}, \eqref{def:fourier} and \eqref{def:transf}, we find
\begin{align}
\begin{split}
    \lim_{q/k\to 0} e^{(s)*}_{ij}S_{ij}   &=
  \sqrt{  \frac{2}{15\pi}}\sum_{s'=0,\pm2}  \xi^{(s')}(\mathbf q)\int \frac{k^2dk}{2\pi^2} \\
 &k^2    
 C_{s',2s} (k,\mathbf q) \left[\tilde A^2(\eta,k)  + \tilde D^2(\eta,k) \right.\\
 &\left.+\frac{4\left(\tilde A(\eta,k) -  \frac{{\tilde D}'(\eta,k)}{\mathcal H(\eta) }\right)^2}{3 (1+w)}\right],
\end{split}
 \label{def:Ker}
\end{align}
where $ C_{s',2s} (k,\mathbf q)$ is the quadrupole harmonic coefficient of $C_{s'} (\mathbf k,\mathbf q)$ for $\hat k$, and we used
\begin{align}
    e^{(\pm)*}_{ij}(\mathbf q)\hat k_i \hat k_j =& \sqrt{\frac{8\pi}{15}}Y^*_{2,\pm2}(\hat k).
\end{align}
Note that we will fix the OPE parameters in the following sections.

To summarize, we find the following superhorizon solution to Eq.~\eqref{eom:tens:source}:
\begin{align}
\begin{split}
    &\lim_{q\to 0}H^{(s)}(\eta,\mathbf q)= \tilde H^{(s)}_{\rm L}(\eta, q)\xi^{(s)}(\mathbf q)\\
    &+\sum_{s'=0,\pm2}\int \frac{k^2dk}{2\pi^2} C_{s',2s}(k,\mathbf q) \mathcal W\left(\eta, k\right)\xi^{(s')}(\mathbf q),
\end{split}
    \label{def:hslm}
\end{align}
where we introduced a $k$ space window function 
\begin{align}
\begin{split}
    \mathcal W\left(\eta,k\right) =&\sqrt{\frac{2}{15\pi}} \int^{\eta}_{\eta_{\rm ini}} d\bar \eta G(\eta,\bar\eta)  \\
    & \times k^2 \left[ \tilde A(\bar \eta, k)^2  +  \tilde D(\bar \eta, k)^2 \right.\\
    &\left.+ \frac{4\left(\tilde A(\bar \eta,k) -  \frac{{\tilde D}'(\bar \eta,k)}{\mathcal H(\bar \eta) }\right)^2}{3 (1+w)}\right].
\end{split}
     \label{w:1}
\end{align}
Thus, $C_{s',2s}(k,\mathbf q)\neq 0$ is necessary to get non-vanishing superhorizon induced tensor perturbations.
$q\to 0$ has two meanings; $q\eta\to 0$ and $q/k\to 0$.
The former means we consider the superhorizon induced perturbations, and the latter implies a hierarchy between the scale of the induced tensor modes and that of the quadratic sources.
Interestingly we may write the nonlinear part of Eq.~\eqref{def:hslm} as
\begin{align}
    \lim_{q/k\to 0}H^{(s)}_{\rm NL}(\eta,\mathbf q)     =& 
    \sum_{s'=0,\pm2}\tilde H^{(s)}_{{\rm NL},s'}(\eta,\mathbf q) \xi^{(s')}(\mathbf q).\label{xi:skemat:2}
\end{align}
Thus, we can treat the induced tensor perturbations as if they are linear perturbations in $q/k\to 0$ limit even when $q\eta$ is not small.
Therefore, we may linearly extrapolate the evolution of the induced tensor perturbations after the horizon entry for $q/k\to 0$ configurations.
This property is useful because we can avoid a complicated calculation of second-order perturbation theory for the evolution after the horizon entry.

\subsection{Induced powerspectrum}
\label{ind:power}

Our observables are B-mode polarization, which is produced by the induced tensor perturbations at the last scattering surface.
The maximum observable wavenumber is the Silk damping scale $k_{\rm D}$ at recombination time $\eta_{\rm rec.}$.
Therefore, we evaluate the induced tensor perturbations of~\footnote{We use a letter of ``$k$'' for a loop momentum and ``$q$'' for an external leg in this article unless otherwise stated.}
\begin{align}
    q<k_{\rm D}(\eta_{\rm rec}).\label{scaleObs}
\end{align}
Note that the solution is written in the form of Eq.~\eqref{xi:skemat:2} for $k>k_{\rm D}(\eta_{\rm rec})$, so we chose the lower bound of $k$ integral \eqref{def:hslm} as $k=k_{\rm D}(\eta_{\rm rec})$ and ignore $k_{\rm D}(\eta_{\rm rec})>k$ modes. 
Also, we evaluate the induced powerspectrum when this mode enters the horizon, that is, at $\eta_{\rm D}\equiv k^{-1}_{\rm D}(\eta_{\rm rec.})$. 
One can justify this prescription because we already know non-Gaussianity is tiny at $k_{\rm D}(\eta_{\rm rec})>k$ through the CMB anisotropy measurements.

\medskip
Now, we define the powerspectrum of the induced tensor perturbations as
\begin{align}
    \begin{split}
      \lim_{q/k\ll 1}&\sum_{s} \langle 
H^{(s)}(\eta_{\rm D},\mathbf q)H^{(s)*}(\eta_{\rm D},\mathbf q')\rangle \\
= &(2\pi)^3\delta^{(3)}(\mathbf q-\mathbf q')P_H (\mathbf q),        
    \end{split}
\label{def:pow:tens}
\end{align}
which can be decomposed into
\begin{align}
    P_H (\mathbf q) =  P_{H,{\rm NL}}(\mathbf q)+ 2 P_{H,{\rm NL}-L}(\mathbf q) + P_{H,{\rm L}}(q).
\end{align}
$P_{H,{\rm NL}}$, $P_{H,{\rm NL}-L}$ and $P_{H,{\rm L}}$ are auto powerspectrum of $H_{\rm NL}$, cross powerspectrum of $H_{\rm NL}$ and $H_{\rm L}$, and auto powerspectrum of $H_{\rm L}$, resepctively.
Note that the induced tensor powerspectrum can be angular dependent: we will show in the following sections that statistically anisotropic non-Gaussianity inevitably induce the statistically anisotropic tensor perturbations.
In the present normalization condition, we can also write $\lim_{q/k\ll 1}\langle H_{ij}(\eta_{\rm D},\mathbf q)H^*_{ij}(\eta_{\rm D},\mathbf q')\rangle = (2\pi)^3\delta^{(3)}(\mathbf q-\mathbf q')P_H (\mathbf q)$.
Therefore, we introduce the tensor-to-scalar ratio as
\begin{align}
    r\equiv \frac{P_H }{P_{\zeta}},
\end{align}
where we introduced the scalar powerspectrum 
\begin{align}
    \langle \zeta(\mathbf q)\zeta(\mathbf q')\rangle = (2\pi)^3\delta^{(3)}(\mathbf q+\mathbf q')P_\zeta(q).
\end{align}
We also use the dimensionless powerspectrum $\mathcal P_\zeta \equiv k^3 P_\zeta/2\pi^2$ in the following sections.

\section{Initial conditions}\label{initial:cd}

Eq.~\eqref{def:hslm} shows that superhorizon induced tensor perturbations are nonzero for quadrupole OPE coefficients.
Indeed, such an angular dependent coefficient does not appear in the most standard scalar non-Gaussianity, but we will see it is not vanishing for anisotropic non-Gaussianity.

\subsection{Case 1: Statistically anisotropic non-Gaussianity}
Let us consider scalar bispectrum defined as
\begin{align}
    \left\langle \zeta(\mathbf k)\zeta(\mathbf k')\zeta(\mathbf q)\right\rangle = (2\pi)^3 \delta^{(3)}(\mathbf k+\mathbf k'+\mathbf q)B_\zeta(\mathbf k,\mathbf k',\mathbf q).
\end{align}
For standard statistically isotropic perturbations, bispectra can be parameterized by at most 3 parameters as $B_\zeta(k,k',q)$. 
On the other hand, some inflationary model with U(1) gauge fields predicts violation of statistical isotropy, which results in the additional angular dependence of the scalar bispectrum~\cite{Bartolo:2012sd, Shiraishi:2013vja, Shiraishi:2016mok}.
More generally, higher spin fields during inflation leave anisotropy in primordial non-Gaussianity~\cite{Arkani-Hamed:2015bza,Lee:2016vti,Shiraishi:2016mok,Baumann:2017jvh,Bordin:2018pca}.
In this article, let us focus on the most straightforward and concrete example of spin 1.
In the squeezed limit, one can write such a bispectrum in the following form~(see e.g., Ref.~\cite{Shiraishi:2013vja})
\begin{align}
\begin{split}
    &\lim_{k\sim k'\gg q}B_\zeta(\mathbf k,\mathbf k',\mathbf q) = 24 P_\zeta(k)P_\zeta(q)g_\star (k) N(q) \\
    &
    \times \left[
    1 - (\hat k\cdot \hat d)^2 - (\hat  q \cdot \hat d)^2
    +
    (\hat k\cdot \hat d)  (\hat  q \cdot \hat d)
    (\hat k\cdot \hat q) 
    \right],
\end{split}
    \label{bispectrum:SA}
\end{align}
where $\hat d$ is a preferred direction in the presence of a vector field during inflation.
$N(q)\sim 60$ e-folds, and $g_\star(k) $ is a parameter which is related to the U(1) gauge field strength and inflationary potential evaluated when a $k$ mode exits the horizon.
This $g_\star$ controls the magnitude of the anisotropy.

One can write the angular dependent part using the spherical harmonics in the following way:
\begin{align*}
        &1 - (\hat k\cdot \hat d)^2 - (\hat  q \cdot \hat d)^2
    +
    (\hat k\cdot \hat d)  (\hat  q \cdot \hat d)
    (\hat k\cdot \hat q) 
      \\
    =& 1 
    -\left(\frac23P_2(\hat k\cdot \hat d)+\frac13\right)
    -\left(\frac23P_2(\hat q\cdot \hat d)+\frac13\right)\\
    &
    +P_1(\hat k\cdot \hat d)P_1(\hat q\cdot \hat d)P_1(\hat k\cdot \hat q)
    \\
            =& \frac{4\pi}{3}\sqrt{4\pi}Y^*_{00}(\hat k)Y_{00}(\hat q)Y^*_{00}(\hat d) \\
            &
    -\frac23
    \frac{4\pi}{5}\sqrt{4\pi}\sum_{M_1=-2}^2Y^*_{2M_1}(\hat k)Y_{00}(\hat q)Y_{2M_1}(\hat d) \\
    &
    -\frac23
        \frac{4\pi}{5}\sqrt{4\pi}\sum_{M_1=-2}^2Y^*_{00}(\hat k) Y_{2M_1}(\hat q)Y^*_{2M_1}(\hat d)
     \\
    &+
    \left (\frac{4\pi}{3}\right)^3
    \sum_{M_1,M_2,M_3}
    Y_{1M_1}(\hat k)Y_{1M_3}(\hat k)Y^*_{1M_3}(\hat q)\\
    &\times
        Y^*_{1M_2}(\hat q)Y^*_{1M_1}(\hat d)Y_{1M_2}(\hat d).
\end{align*}
Then we can expand the squeezed limit bispectrum as follows: 
\begin{align}
\begin{split}
    &\lim_{k\sim k'\gg q}B(\mathbf k,\mathbf k',\mathbf q) = 24g_\star (k) N(q) P_\zeta(k)P_\zeta(q)
    \\& \times \sum_{\ell_im_i} B^{\ell_1\ell_2\ell_3}_{m_1m_2m_3}Y^*_{\ell_1 m_1}(\hat k)Y_{\ell_2 m_2}(\hat q)Y^*_{\ell_3 m_3}(\hat d),    
\end{split}
\end{align}
where $i$ runs from 1 to 3, and multi-harmonic coefficients are defined as
\begin{align}
\begin{split}
    &B^{\ell_1\ell_2\ell_3}_{m_1m_2m_3}\\
    =& 
    \frac{4\pi}{3}\sqrt{4\pi} \delta_{\ell_1 0}\delta_{m_10}\delta_{\ell_2 0}\delta_{m0}
    \delta_{\ell_3 0}\delta_{m_30}\\
    &
    -\frac23
        \frac{4\pi}{5}\sqrt{4\pi}
        \delta_{\ell_1 2} 
        \delta_{\ell_2 0 }\delta_{m_2 0}
        (-1)^{m_1} \delta_{\ell_3 2}\delta_{m_3,- m_1}
   \\
        &
        -\frac23
    \frac{4\pi}{5} \sqrt{4\pi}  
        \delta_{\ell_1 0}\delta_{m_1 0} 
        \delta_{\ell_2 2}
        \delta_{\ell_3 2}\delta_{m_3,m_2}
        \\
    &+
    \left (\frac{4\pi}{3}\right)^3
    \sum_{M_1,M_2,M_3}
    \mathcal G^{\ell_1,1,1}_{m_1,M_1,M_3}\\
    &\times
   \left(    \mathcal G^{\ell_2,1,1}_{m_2,M_2,M_3}\right)^*
    (-1)^{M_1}
    \mathcal G^{\ell_3,1,1}_{m_3,-M_1,M_2}
    , 
\end{split}
\end{align}
with the Gaunt integral
\begin{align}
    \mathcal G^{\ell_1\ell_2\ell_3}_{m_1m_2m_3} \equiv \int d\hat n Y_{\ell_1m_1}(\hat n)Y_{\ell_2m_2}(\hat n)Y_{\ell_3m_3}(\hat n).
\end{align}
We find the OPE coefficient for $s'=0$ as
\begin{align}
\begin{split}
    C_0(\mathbf k,\mathbf q)= & 24g_\star (k) N(q) P_\zeta(k)\sum_{\ell_im_i} B^{\ell_1\ell_2\ell_3}_{m_1m_2m_3}\\ &\times Y^*_{\ell_1 m_1}(\hat k)Y_{\ell_2 m_2}(\hat q)Y^*_{\ell_3 m_3}(\hat d).
\end{split}
\end{align}
Then, Eq.~\eqref{xi:skemat:2} becomes
\begin{align}
\begin{split}
    &H^{(s)}_{\rm NL}  = \zeta h_0 \sum_{\ell_im_i}B^{2,\ell_2\ell_3}_{-s,m_2m_3}  Y_{\ell_2 m_2}(\hat q)Y^*_{\ell_3 m_3}(\hat d),
\end{split}
\end{align}
where we have defined
\begin{align}
    h_0(\eta,q)\equiv \int \frac{k^2dk}{2\pi^2}P_\zeta(k) 24g_\star (k) N(q)   \mathcal W\left(\eta, k\right).\label{def:h0}
\end{align}
Without loss of generality, we may choose a $z$ axis, so let us consider $\hat z\parallel \hat q$.
In this case, we find
\begin{align}
    Y_{\ell m}(\hat q=\hat z)=\delta_{m0}\sqrt{\frac{2\ell+1}{4\pi}}.
\end{align}
Then, we may simplify $r$ by using Legendre polynomials:
\begin{align}
\begin{split}
r
    =&
    \sum_{s,\ell,\ell_2 \ell_3
    \ell'_2 \ell'_3}
    \sqrt{\frac{2\ell+1}{4\pi}}
    \sqrt{\frac{2\ell_2+1}{4\pi}}\sqrt{\frac{2\ell'_2+1}{4\pi}}\\
    &\times 
    B^{2,\ell_2\ell_3}_{-s,0,s}B^{2,\ell'_2\ell'_3*}_{-s,0,s} 
    \mathcal G^{\ell\ell_3\ell_3'}_{0,-s,s}
    P_{\ell }(\hat d\cdot \hat q) 
    h_0^2,
\end{split}
    \label{Expand:Led}
\end{align}
where we assume that the primary tensor perturbations are subdominant for simplicity. 
Simplifying Eq.~\eqref{Expand:Led}, we get
\begin{align}
\begin{split}
     r=&\frac{8h_0^2 }{225} \left[1   -\frac{10}{7}   P_2\left(\hat q\cdot \hat d\right)+\frac{3}{7}  P_4\left(\hat q\cdot \hat d\right)\right] .
\end{split}
\end{align}
Introducing Legendre expansion of the tensor-to-scalar ratio
\begin{align}
    r = \sum_{\ell=0}^{4}(-i)^\ell(2\ell+1)r_\ell P_\ell \left(\hat q\cdot \hat d\right),\label{expand:Le}
\end{align}
we find the monopole component
\begin{align}
    r_0 = \frac{8h_0^2 }{225}.\label{SA:STR}
\end{align}
$r$ depends on $q$ through $N(q)$, which is almost scale invariant at the CMB scale. 
Thus, we can have a scale-invariant tensor powerspectrum without primordial tensor perturbations.
Moreover, the tensor powerspectrum becomes statistically anisotropic, and we obtain the following consistency relations:
\begin{align}
    \frac{r_2}{r_0}=\frac{2}{7},~\frac{r_4}{r_0}=\frac{1}{21}, ~\frac{r_4}{r_2}=\frac{1}{6},\label{Ota:cons}
\end{align}
which can be useful to distinguish the induced tensor powerspectrum from the other initial conditions.

\subsection{Case 2: scalar-scalar-tensor non-Gaussianity}
\label{case2:sst}
Another possibility of angular dependent OPE coefficients is scalar-scalar-tensor non-Gaussianity:
\begin{align}
    \begin{split}
        &\left\langle \zeta(\mathbf k)\zeta(\mathbf k')H^{(s)}(\mathbf q)\right\rangle \\
        = &(2\pi)^3 \delta^{(3)}(\mathbf k+\mathbf k'+\mathbf q)B^{(s)}_{\zeta\zeta H}(\mathbf k,\mathbf k',\mathbf q).
    \end{split}
\end{align}
This type of scalar-tensor cross bispectrum was discussed in Refs.~\cite{Maldacena:2002vr,Seery:2008ax,Shiraishi:2010kd,Gao:2012ib,Franciolini:2017ktv,Domenech:2017kno,Biagetti:2017viz}.
In particular, authors in Ref.~\cite{Domenech:2017kno} showed that massive spin-2 fields potentially enhance this correlation without changing the standard predictions of single-field inflation. 
In the squeezed limit, $B^{(s)}_{\zeta\zeta H}(\mathbf k,\mathbf k',\mathbf q)$ can be written in the following form~\cite{Maldacena:2002vr}
\begin{align}
\begin{split}
    &\lim_{k\sim k'\gg q} B^{(s)}_{\zeta\zeta H}(\mathbf k,\mathbf k',\mathbf q) \\
    = &\frac{3}{4}\alpha^{(s)}(q) P_{H,{\rm L}}(q)\beta(k) P_\zeta(k)  e^{(s)}_{ij}(\mathbf q)\hat k_i \hat k_j,
\end{split}
    \label{tens:NG}
\end{align}
where we assumed separable coefficient $\alpha^{(s)}\beta$, which is $1$ for slow-roll single field inflation.
By definition, primary tensor perturbations are inevitable in this case.
Hence, one would be more interested in a possibility such that the induced effects dominate the primordial tensor perturbations.
Combining Eqs.~\eqref{OPE2} and \eqref{tens:NG}, we find
\begin{align}
    C_{\pm2}(\mathbf k,\mathbf q) = \frac{3}{4}\alpha^{(\pm2)}(q) \beta(k) P_\zeta(k)
\sqrt{\frac{8\pi}{15}}Y_{2,\pm2}(\hat k).
\end{align}
Then $s'=\pm2$ coefficient in Eq.~\eqref{xi:skemat:2} is nonzero, and we obtain 
\begin{align}
    H^{(s)}_{\rm NL}
=& h_{s}\xi^{(s)},
\end{align}
where we have defined
\begin{align}
    h_{\pm2}(\eta,q)\equiv \sqrt{\frac{3\pi}{10}} \int \frac{k^2dk}{2\pi^2}\alpha^{(\pm2)}(q) \beta(k) P_\zeta(k)
 \mathcal W\left(\eta, k\right).
\end{align}
The total tensor powerspectrum is
\begin{align}
    P_H  =  \frac{P_{H,{\rm L}}}{2}\sum_{s=\pm2}(1+h_s)^2.\label{PHtss}
\end{align}
Thus, we cannot distinguish the induced powerspectrum from the primary tensor powerspectrum as long as $\alpha$ is scale invariant. 
Let us write the cross term of the linear and nonlinear parts of the tensor-to-scalar ratio as 
\begin{align}
    r_{\rm NL- L} \equiv  \frac{2 P_{H,{\rm NL}-L}}{P_{H,{\rm L}}} .\label{rnl/r}
\end{align}
If $|r_{\rm NL- L}| > 1$, the observed tensor perturbations are dominated by the induced contributions.
Therefore, $r_{\rm NL}$ can be large enough to observe even if the primary contribution is small.

\section{Evaluation of the induced tensor perturbations}
\label{sec:window}

In the previous sections, we presented a theoretical framework of the superhorizon induced tensor perturbations.
The first goal of this article is to evaluate Eqs.~\eqref{SA:STR} and \eqref{rnl/r} by computing the window function~\eqref{w:1} and integrating it in $k$ space.

\subsection{Characteristic scales}

As we discussed in Eq.~\eqref{def:pow:tens}, we evaluate the superhorizon tensor powerspectrum at some conformal time before horizon entry of the Silk damping scale at recombination.
In this paper, we chose $\eta_{\rm D}=10.$Mpc in Eq.~\eqref{def:pow:tens}.
Then the lower limit of the $k$ space integral \eqref{def:hslm} is $k_{\rm D}=0.1{\rm Mpc}^{-1}$ since $\mathcal H =1/\eta$ in the radiation era.
On the other hand, the upper bound of $k$ space integral should be the horizon scale $k_{I}$ at the end of inflation.
The present Hubble parameter is $67{\rm km \cdot s}^{-1}\cdot {\rm Mpc}^{-1}$, so we get $k_0=2.2\times 10^{-4}{\rm Mpc}^{-1}$.
Then $k_{I}= \mathcal H(\eta_I) \mathcal H(\eta_0)^{-1} k_0=e^{60}k_0=2.6\times 10^{22}{\rm Mpc}^{-1}$ for inflation of e-fold 60.
Therefore, we integrate the window function for $0.1\lesssim k {\rm Mpc} \lesssim 10^{22}$ and evaluate it at $\eta=10.{\rm Mpc}$.

The Universe has experienced different regimes before recombination since inflation terminated.
First of all, if the inflaton or some other fields oscillate at the bottom of inflationary potential, the Universe becomes matter dominant effectively.
We call this epoch an early matter-dominated epoch.
Depending on preheating or reheating scenarios, this early matter era takes a short time or a long time.
Then, fields decay into radiation, reheating terminates, and the Universe becomes radiation dominant.
Finally, cold darkmatter starts to dominate after redshift $z\sim 3400$, or $k_{\rm eq}=2(2-\sqrt{2})\eta^{-1}_{\rm eq}\sim 0.01{\rm Mpc}^{-1}$.
We call this period the late matter era.

The evolution of the induced tensor modes highly depends on the evolution of the background Universe, or more directly, the evolution of the gravitational potential.
In the radiation era, the gravitational potential is decaying after the horizon entry, and fluid velocity is not growing.
As a result, the secondary source is damping soon after the horizon entry.
On the other hand, during the matter era, the linear gravitational potential is constant on all scales, and the fluid velocity is growing on small scales.
Therefore, the source is not decaying, and tensor perturbations can be induced a lot in the early matter era.

In contrast to the transition from the radiation era to the late matter era, we know very little about the transition from the early matter era to the radiation era.
Even worse, we have no idea when it happened.
Hence, reheating temperature $T_R$ or reheating conformal time $\eta_R\sim k^{-1}_{R}$ are free parameters in the following analysis.
$\eta_I\sim k_I^{-1}$ is the minimum possible value of $\eta_R$ for instantaneous reheating after inflation. 
In most cases, we assume $T_R$ should be above the temperature of Big Bang nucleosynthesis~(BBN).
    The relation between $T_R$ and $\eta_R$ is given as~\cite{Inomata:2019ivs}
\begin{align}
 \begin{split}
     \eta_R &= 10^{-14}{\rm Mpc}\left(\frac{g_s}{106.75}\right)^{1/3}\left(\frac{g}{106.75}\right)^{-1/2}\\
     &\times \left(\frac{T_{\rm R}}{1.2\times 10^7{\rm GeV}}\right)^{-1},
 \end{split}   
\end{align}    
with the effective degrees of freedom for the entropy density $g_s$ and the effective relativistic degrees of freedom $g$.
For simplicity, we ignore the details of transition; we evaluate contributions from the radiation era and early matter era separately.
However, Refs.~\cite{Inomata:2019ivs, Inomata:2019zqy} studied induced tensor perturbations for a sudden transition and a gradual transition from the early matter era to the radiation era, providing concrete models.
In this paper, we will comment on the superhorizon induced spectrum for the sudden transition after some conservative analysis.

\subsection{Analytic transfer function of the gravitational potential}
\label{anal;transfsec}

In this article, we do not integrate the Einstein equation full numerically.
Instead, we use analytic transfer functions of the gravitational potential with several assumptions.
First, when we ignore the anisotropic stress in the Einstein equation, one finds $A=-D$.
Note that this assumption is not valid deep inside the horizon.
Then, we assume the equation of state is constant.
For an arbitrary constant $w$, with $a\propto \eta^{\frac{2}{1+3w}}$, we find~\cite{Mukhanov:2005sc}
\begin{align}
   \tilde  A'' + \frac{6(1+w)}{1+3w}\frac{1}{\eta}\tilde A' + w k^2 \tilde A =0,
\end{align}
and its general solution
\begin{align}
    \tilde A = (\sqrt{w}k\eta)^{-\nu} \left[
    C_1  J_\nu(\sqrt{w}k\eta)+C_2 Y_\nu(\sqrt{w}k\eta)
    \right],\label{gen:sol}
\end{align}
where $\nu \equiv (5+3w)/2(1+3w)$.
$J_\nu$ and $Y_{\nu}$ are the Bessel functions of the first kind and the second kind, respectively.
In each era, matching the coefficients on superhorizon $k\eta \to 0$, we obtain
\begin{align}
    \tilde A = \frac{3+3w}{5+5 w}\Gamma[\nu+1]\left(\frac{\sqrt{w}k\eta}{2}\right)^{-\nu}J_\nu(\sqrt{w}k\eta),
\end{align}
where $\Gamma$ is the Gamma function.
For $w=1/3$, one finds~\cite{Dodelson:2003ft}
\begin{align}
    \tilde A(\eta,k) = \frac{2}{3}\frac{9}{k^2\eta^2}\left(\frac{\sin(k\eta/\sqrt3)}{k\eta/\sqrt3}-\cos(k\eta/\sqrt3)\right).
\end{align}
Also, for $w=0$, we get $\tilde A=-\tilde D=3/5$ on superhorizon scale.
In the early matter era, we write the transfer functions as 
\begin{align}
    \tilde A(\eta,k)= \frac{3}{5}\theta(k_{\rm cut}(\eta)-k),\label{eq:A:anal:eMD}
\end{align}
where we introduced the step function $\theta$ to include the cut-off scale.
Two factors determine this cut-off scale.
First, the early matter era starts when inflation ends at $\eta=\eta_I$.
The initial condition of the gravitational potential is well-defined on the superhorizon scale at that time, so we drop modes inside the horizon at $\eta_I$ for simplicity.
Second, in the matter-dominated epoch, density perturbations grow nonlinearly.
Using the Poisson equation 
\begin{align}
    2M_{\rm pl}^2 k^2 A =a^2\rho \delta     
\end{align}
and the Friedmann equation
\begin{align}
    3\mathcal H^2 M_{\rm pl}^2=a^2 \rho,
\end{align}
with $\mathcal H =2/\eta$ in matter era, we find 
\begin{align}
    A = \frac{6}{k^2\eta^2} \delta. 
\end{align}
As $A$ is constant in matter era, and we have $A=3/5\sqrt{\mathcal P_\zeta}$ for the scale invariant curvature perturbations.
Assuming $\mathcal P_\zeta=2.2 \times 10^{-9}$, we find $\delta=1$ for $k\eta \sim 462$.
Solving this equation, the authors in Ref.~\cite{Assadullahi:2009nf,Alabidi:2013wtp,Inomata:2019ivs} concluded that the non-linear cut-off scale is given by $462/\eta$ and hence
\begin{align}
    k_{\rm cut}={\rm min.}[462/\eta,1/\eta_{I}].\label{cut:1}
\end{align}
However, nonlinear evolution normally starts before $\delta$ gets unity in linear theory.
So, writing the full density contrast as $\delta=\delta_{\rm L}+\delta_{\rm NL}$, we should get $\delta_{\rm NL}=1\gg\delta_{\rm L}$.
Thus, Eq.~\eqref{cut:1} does not provide a cut-off scale of linear perturbation theory properly.
In other words, when we get $\delta_{\rm L}=1$, one finds the full density contrast is above unity. 
Nevertheless, we may justify computing tensor perturbations using Eq.~\eqref{cut:1} in the present case.
This justification is because different OPEs coefficients determine the significance of the nonlinear terms.
For example, one can parameterize a cubic order operator product by the conventional local form cubic order parameter $g_{\rm NL}$.
However, such a contribution is normally subdominant because higher-order non-Gaussianity employs an additional $\mathcal P_\zeta$.
Also, the gravitational potential is not highly nonlinear even if the matter perturbations become large.
Therefore, the quadratic term of linear perturbations would be dominant for the source of superhorizon tensor perturbations, even in the nonlinear regime.
Then, can we extend the cut-off scale to an arbitrarily small scale?
The answer is ``no'', but we point out that one can extend the cut-off scale more.
In this case, the crucial scale is given by the critical density $\delta_{\rm c}=1.69$ rather than $\delta=1$.
In the Press-Schechter theory for halo formation, a spherical region collapses when the linear density perturbations reach the critical density~(see, e.g., \cite{Mukhanov:2005sc}, for a review of the nonlinear collapse).
One usually considers that shell crossing happens at some point in the non-perturbative regime, and dense regions are virialized.
We account for this smoothing effect by introducing the cut-off scale with a step function~\eqref{eq:A:anal:eMD}; therefore, more optimistically, we may improve the cut-off scale as 
\begin{align}
    k_{\rm cut}={\rm min.}[585/\eta,1/\eta_{I}].
\end{align}
In this article we will use this cut-off scale instead of Eq.~\eqref{cut:1}.

\subsection{Window functions}

\begin{figure}
\centering
  \includegraphics[width=\linewidth]{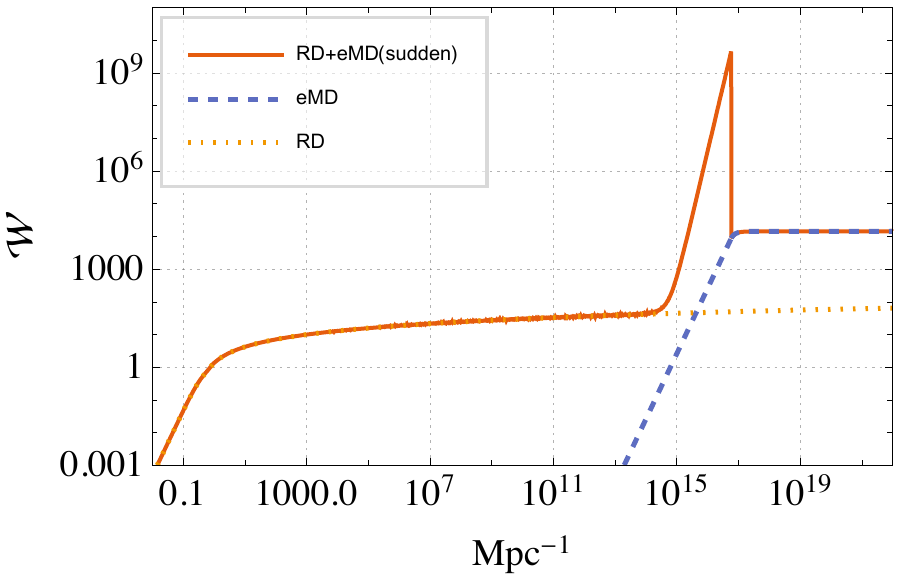}
  \caption{Plots of the window functions for the early matter era, the radiation era and their sudden transition. We chose $T_{\rm R}=1.2\times 10^7$GeV, which is equivalent to $\eta_R=10^{-14}$Mpc.
  The secondary source for radiation era~(dashed line) is almost scale-invariant, and the window for early matter era~(dot-dashed line) is $10^3$ times bigger than that of the radiation era.
  The solid orange line accounts for both era and their sudden transition.
  }
  \label{w:fig:MD:RD}
\end{figure}

Using the above approximate formulae, let us evaluate the window function~\eqref{w:1} at $\eta=\eta_{\rm D}$.
We show plots of the window function~\eqref{w:1} in Fig.~\ref{w:fig:MD:RD}.
In Fig.~\ref{w:fig:MD:RD}, we evaluated the window function for the radiation era and the early matter era with $T_{R}=1.2\times 10^7$GeV.
The contribution from the radiation era is asymptotically flat on high $k$.
We can see the window function is enhanced for the early matter era, as was pointed out in Ref.~\cite{Assadullahi:2009nf}. 
The plateau during early matter era is because of the step function $\theta(k_{\rm cut}(\eta)-k)$. 
If we do not have this step function, the window function blows up because of the linearly extrapolated high $k$ modes. 

Refs.~\cite{Inomata:2019ivs, Inomata:2019zqy} showed that details of the transition from the early matter era to the radiation era might affect the induced tensor perturbations.
They found that, depending on the transition time scale, induced tensor perturbations can be amplified or suppressed significantly.
For example, if it is a sudden transition, scalar perturbations deep inside the horizon at the initial time of the radiation era strongly enhance the induced spectrum.
This enhancement is because high $k$ modes oscillate quite rapidly after the transition.  
We find the scalar transfer functions by matching the gravitational potential and its derivative at $\eta=\eta_R$ using Eq.~\eqref{gen:sol}, following Ref.~\cite{Inomata:2019ivs}.
Note that they pointed out we may assume continuity of the gravitational potential on subhorizon scales, while it is discontinuous on superhorizon scales.
Then, in Fig.~\ref{w:fig:MD:RD}, we showed the enhancement due to the sudden transition from the early matter era to the radiation era.
We can see a huge enhancement in $k_R<k<k_{\rm cut}(\eta_R)$.
The peak of the window function is $\mathcal O(10^{9})$ at $k=k_{\rm cut}$.
We can reproduce this value by evaluating Eq.~\eqref{w:1} as
\begin{align}
    \mathcal W(\eta,k_{\rm cut}) \sim \sqrt{\frac{2}{15\pi}}\frac{(k_{\rm cut}\eta)^4}{4}\frac{4}{3(1+w)}\left(\frac{3}{5}\right)^2 \Bigg|_{w=\frac13}.\label{Wsdaprox}
\end{align}
Thus, the term proportional to $(D'/\mathcal H)^2$ in Eq.~\eqref{w:1} gives the peak.

\section{Tensor-to-scalar ratio for statistical anisotropy}
\label{tsr:sa}
Now we discuss observability of the induced tensor powerspectrum.
In this section, we estimate the tensor-to-scalar ratio for the statistically anisotropic scalar non-Gaussianity~\eqref{bispectrum:SA}.
Here, we assume $g_\star$ is scale-invariant for simplicity.

\subsection{Induced spectrum in the radiation era}
\label{cons:anal}

\begin{figure}
    \centering
  \includegraphics[width=\linewidth]{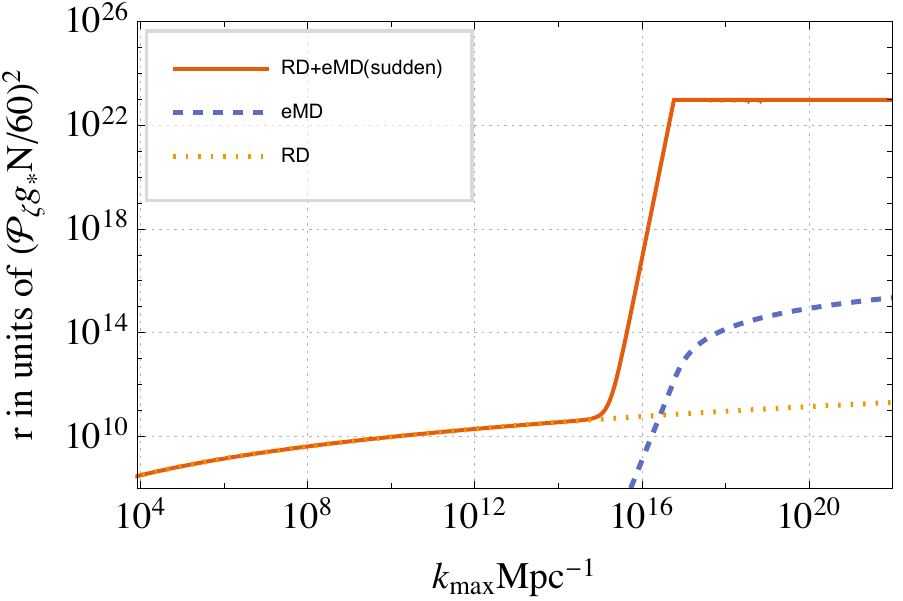}
  \caption{Tensor-to-scalar ratio as a function of $k_{\rm max}$ for the early matter era, the radiation era and their sudden transition. We chose $T_{\rm R}=1.2\times 10^7$GeV, which is equivalent to $\eta_R=10^{-14}$Mpc.
  The logarithmic divergence is slow enough for practical values of $k_{\rm max}$.
  }
  \label{fig:r:MD:RD}
\end{figure}

As we mentioned, $T_R$ is a model-dependent parameter in the present analysis, and the choice of $T_R$ changes estimations drastically.
Therefore, we first give the most conservative estimation of the spectrum in the radiation era and discuss upper bounds on parameters.
In Fig.~\ref{fig:r:MD:RD}, we plot Eq.~\eqref{SA:STR} as a function of $k_{\rm max}$, which is the integral upper bound in Eq.~\eqref{def:hslm}.
For pure radiation era and $k_{\rm max}{\rm Mpc}=10^{22}$, Fig.~\ref{fig:r:MD:RD} shows
\begin{align}
    r_0 \sim 10^{11}(\mathcal P_\zeta g_\star)^2\label{r:in:zeta:gstar}.
\end{align}
Since the $k$ space window function is nearly constant as we showed in~Fig.~\ref{w:fig:MD:RD}, we see $r$ is logarithmically divergent but is not sensitive to the practical choice of $k_{\rm max}$.

\begin{figure}
\centering
  \includegraphics[width=\linewidth]{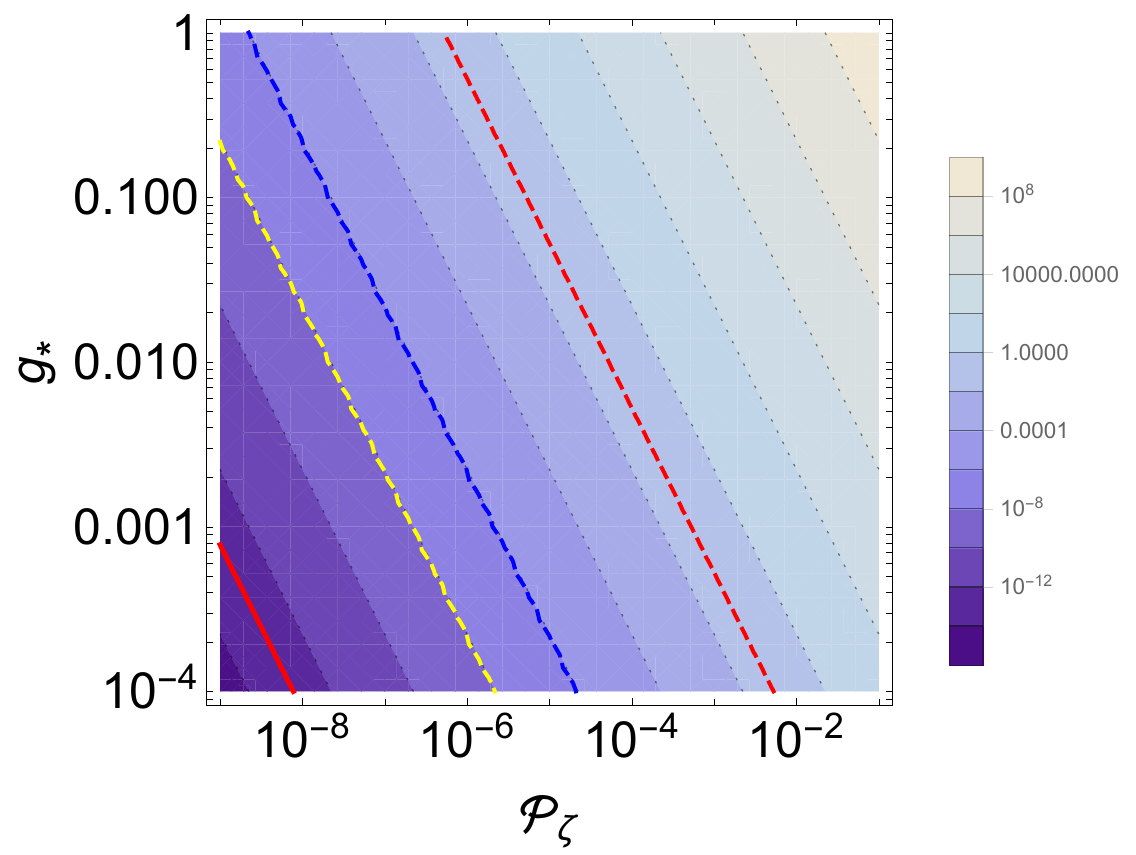}
  \caption{A contour plot of the (monopole) tensor-to-scalar ratio induced in the radiation era.
  The red dashed line corresponds to the current upper bound from the observations of the CMB polarization.
  The blue dashed line corresponds to $r=10^{-6}$, which is comparable to noise due to the induced gravitational waves from Gaussian initial conditions.
  The yellow dashed line means $r=10^{-8}$. This line gives the observable lower limit of anisotropic part of $r$ for the above noise.
  The red solid line includes the enhancement due to sudden transition from the early matter era to the radiation era.
  }
  \label{contourplot}
\end{figure}

Now, let us discuss upper bounds on statistical anisotropy.
No one knows about short-wavelength density powerspectrum, so we constrain or forecast a pair of $(\mathcal P_\zeta,g_\star)$.
Fig.~\ref{contourplot} is a contour plot of tensor-to-scalar ratio in $(\mathcal P_\zeta, g_\star)$ plane.
The red dashed line corresponds to $r=0.056$, so the current observations exclude the parameter region above this line~\cite{Akrami:2018odb}.
The interval of the contour is 100 times, as shown in the legend.
For the scale-invariant scalar powerspectrum, the upper bounds on $g_\star$ is much weaker than unity.
However, once we have enhancements of $\mathcal P_\zeta$ or $g_\star$ in $k_{\rm D}<k<k_{\rm max}$, the signal would be detectable.
The blue and yellow dashed line corresponds to $r=10^{-6}$ and $r=10^{-8}$, respectively.

\medskip
In Ref.~\cite{Mollerach:2003nq}, the authors estimated the B-mode powerspectrum due to the induced tensor perturbations.
In the previous analysis, the initial condition was Gaussian so that the scalar perturbations on $k_{\rm rec.}<k<k_{\rm D}$ sourced the induced tensor powerspectrum, as we discussed in section~\ref{cmb:mudistortion}.
They showed that this signal corresponds to $r\sim 10^{-6}$, which can be noise for the monopole tensor-to-scalar ratio unless we observe primordial tensor perturbations of $r>10^{-6}$.
Also, as Ref.~\cite{Seljak:2003pn} has discussed possible lower bounds of detectable $r$, we can remove lensing contaminations to observe $r\sim 10^{-6}$.
Then, once we measure tensor-to-scalar ratio $r_{\rm obs}=\max[r_{\rm L},r_{\rm NL},10^{-6}]$, what is the measurable lowest value of $g_{\star}$? 
For the monopole tensor-to-scalar ratio, apparently we cannot see the signal below $r_{\rm obs}$, so the error bar is $\Delta r_0 \sim  r_{\rm obs}$, which corresponds to $\Delta g_\star \sim 20$ for $r_{\rm obs}= 10^{-6}$.
On the other hand, as we showed in Eq.~\eqref{Ota:cons}, the non-Gaussian induced contribution is statistically anisotropic and has $P_2$ and $P_4$ dependence.
Consistency relation \eqref{Ota:cons} is useful to test anisotropic initial conditions.
Ref.~\cite{Hiramatsu:2018vfw} studied detectability of statistical tensor anisotropy.
They showed that we might observe the Legendre coefficients of the tensor powerspectrum up to 1\% level of the observed monopole component~\footnote{Note that their definitions of the Legendre coefficients are different from Eq.~\eqref{expand:Le}}.
Therefore, the error bar would be $\Delta r_2,~\Delta r_4\sim r_{\rm obs}/100$.
Suppose we get $r_{\rm obs}=10^{-6}$, $\Delta r_2\sim \Delta r_4\sim 10^{-8}$. We showed $10^{-8}$ with the yellow line in Fig.~\ref{contourplot}, which corresponds to $\Delta g_\star=0.2$.
Thus, for the most conservative setup, the error bar on $g_\star$ is not improved, compared to $\Delta g_\star =0.02$, which we observed in the previous analysis with CMB temperature anisotropies~\cite{Kim:2013gka}.
However, our forecast covers complementary scales for the previous analysis based on Planck.
They derived the strong upper bound on $g_\star(k)$ for $k_0<k<k_{\rm D}$.
On the other hand, the present induced tensor perturbations are sensitive to $g_\star(k)$ on $k_{\rm D}<k<k_{I}$.
Thus, they studied $g_{\star}$ of 6 to 7 e-folds only, but our forecast covers the rest of e-folds more than 50.

\subsection{Induced spectrum in the early matter era}

\begin{figure}
\centering
  \includegraphics[width=\linewidth]{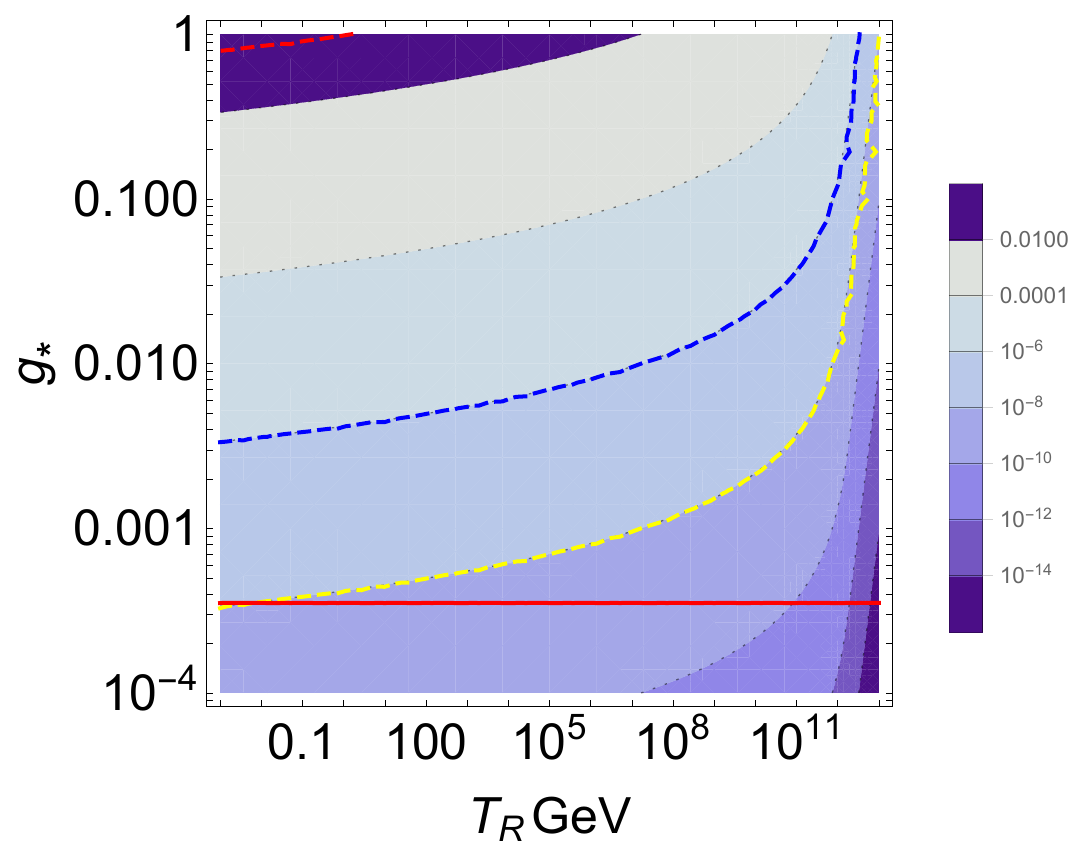}
  \caption{A contour plot of the (monopole) tensor-to-scalar ratio induced in the early matter era in $T_R$-$g_\star$ plane.
 We define the dashed lines in the same way as Fig.~\ref{contourplot}.
  The solid red line includes the enhancement due to sudden transition from early matter era to radiation era.}
  \label{contourplot:TR}
\end{figure}

Next, we consider the tensor-to-scalar ratio for various reheating temperatures.
In Fig.~\ref{fig:r:MD:RD}, we can see $r$ is enhanced during the early matter era by a factor of $10^3$ to $10^4$, depending on the reheating temperature.
Assuming scale invariant curvature perturbations, i.e, $\mathcal P_\zeta=2.2\times 10^{-9}$, we show a contour plot of $r$ in $T_{\rm R}$-$g_\star$ plane in Fig.~\ref{contourplot:TR}.
Then, $\Delta g_{\star}$ improves by a factor of $10^3$ to $10^4$.
Thus, if there exists an early matter era, it has a considerable impact on the superhorizon induced tensor modes.

\subsection{Enhancement due to sudden transitions}

We also leave a comment on the enhancement of the induced tensor perturbations for the sudden transition from the early matter era to the radiation era.
The tensor-to-scalar ratio as a function of $k_{\rm max}$ is shown in Fig.~\ref{fig:r:MD:RD}.
As we see in this plot, the enhancement due to the transition is dominant, and hence $T_R$ dependence is lost.
We found that the tensor-to-scalar ratio is amplified by a factor of $\mathcal O(10^{12})$ compared to the pure radiation case, which we can estimate as
\begin{align}
    \frac{\frac{8}{225} \cdot \left[24\cdot 60\cdot \frac14 \mathcal W(\eta,k_{\rm cut})\right]^2}{10^{11}} \sim 10^{12},    
\end{align} 
where we used Eqs.~\eqref{def:h0}, \eqref{SA:STR} and \eqref{Wsdaprox}.
In Figs.~\ref{contourplot} and \ref{contourplot:TR}, we also show this enhancement with a solid red line. 
Thus, in this case, the present upper bound on $r$ has already strongly constrained the anisotropy as $g_{\star}<3.5\times 10^{-4}$.

\section{Tensor-to-scalar ratio for scalsr-scalar-tensor non-Gaussianity}\label{tsr:sst}

In Eq.~\eqref{PHtss}, we showed that the induced powerspectrum and the primary spectrum scale in the same way.
Also, Eq.~\eqref{Ota:cons} is not applicable for scalar-scalar-tensor non-Gaussianity.
Thus, in principle, we cannot distinguish these tensor spectra, which would be a problem if we observe the intrinsic B-mode polarization in the future.
Of course, the nonlinear part is dominant only for large parameters
\begin{align}
    |\alpha \beta| \times  \left(\frac{\mathcal P_\zeta}{2.2 \times 10^{-9}} \right) \gtrsim 1.4\times 10^{5},
\end{align}
when we only account for the radiation era~(we assume $\alpha^{(\pm2)}=\alpha$ and $\beta$ are scale invariant.).
However, such a large parameter is not so ridiculous if we consider primordial black hole formation in the early Universe. 
Let us consider the scalar powerspectrum has a peak at $k=k_p$.
Here, we model this powerspectrum using the delta function as~\cite{Saito:2008jc}
\begin{align}
    \mathcal P_{\zeta}(k) = \mathcal A^2 \delta (\ln k/ k_p),
\end{align}
where typically we have $\mathcal A^2\sim \mathcal P_\zeta (k_p)\times $(peak width).
Using this spectrum, we find
\begin{align}
 \frac{r_{\rm NL- L}}{r_{\rm L}} \sim 2\sqrt{\frac{6\pi}{5}}\mathcal A^2  \alpha(q) \beta(k_p)  \mathcal W(\eta_{\rm D}, k_{\rm p}).
\end{align}
While we have to specify an inflationary model to fix $|\alpha \beta|$, the induced tensor mode can be dominant for $\mathcal A^2=\mathcal O(10^{-2})$ and $|\alpha \beta|=\mathcal O(1)$.
On the other hand, even for the scale-invariant perturbations with small nonlinear parameters, the induced spectrum can be enhanced, depending on reheating scenarios.
For sudden transition in the previous section, we find
\begin{align}
    \frac{r_{\rm NL- L}}{r_{\rm L}} \sim 4.8 \times \alpha \beta \times \left(\frac{\mathcal P_\zeta}{2.2 \times 10^{-9}} \right).
\end{align}
Thus, even for $\alpha\beta=1$, tensor perturbations are dominated by the induced contribution.
Note that any detection of tensor powerspectrum suffers from this degeneracy unless we exclude the above nonstandard early Universe physics.

\section{Conclusions}

In this paper, we studied induced tensor perturbations, at second-order in scalar perturbations, sourced by mode coupling effects originated from primordial non-Gaussianity.
We derived a superhorizon solution of the induced tensor perturbations and computed the source term, applying OPE for cosmological perturbations.
The amplitude of the induced spectrum is sensitive to the size of anisotropic non-Gaussianity and scalar perturbations on all scales from the horizon scale at the end of inflation to the Silk damping scale in the CMB anisotropies.
Thus, the induced powerspectrum can be a powerful tool to test small scale physics, including primordial blackhole formations and reheating scenarios in the early Universe.
We also found that the induced superhorizon tensor powerspectrum becomes scale-invariant in the presence of anisotropic non-Gaussianity; therefore, the on-going measurements of CMB polarization B-modes are more useful to detect these effects rather than the laser interferometers.
We evaluated the induced tensor-to-scalar ratio for two concrete examples of anisotropic non-Gaussianity: statistically anisotropic scalar non-Gaussianity motivated by some inflationary model with a vector field, and scalar-scalar-tensor non-Gaussianity, which is more common for various inflationary models.

In Fig.~\ref{contourplot}, we presented a contour plot of the tensor-to-scalar ratio for a given scalar powerspectrum and statistical anisotropy.
Upper bounds on $g_\star $ derived from this plot are much weaker than the present constraint by Planck.
However, Planck only tested large scale isotropies on 6-7 e-folds, while our bounds apply to the rest of the total e-folds.
Thus, we showed that the induced tensor powerspectrum offers a test of statistical anisotropy on smaller scales.
In other words, if there are large anisotropies on tiny scales, the induced contribution contaminates the primordial tensor powerspectrum.
This degeneracy is potentially a problem if we observe the primary B-modes in the future because we cannot conclude the energy scale of inflation without excluding the possibility of observing the secondary spectrum.
For the statistically anisotropic non-Gaussianity, the induced powerspectrum also becomes statistically anisotropic, so that we can distinguish the signals by looking at the angular dependence.
On the other hand, the induced spectrum from the scalar-scalar-tensor non-Gaussianity completely degenerates with primordial contributions.
This signal can be dominant only for considerable scalar-scalar-tensor non-Gaussianity.
However, we showed that the induced signal could increase in the case of a specific reheating scenario with a sudden transition from the early matter era to the radiation era.
Similarly, primordial blackhole formations lead to a significant enhancement of the induced tensor modes. 
Therefore, it is essential to specify the early Universe physics to identify the origin of the tensor powerspectrum if we measure it in the future.
In this sense, combining the CMB polarization observations with gravitational wave laser interferometer experiments would be crucial.

\medskip
Lastly, we point out future directions related to the formalism presented here.
First, in this paper, we did not consider the details of gauge dependence of the second-order tensor perturbations, but this would be important for the present case because we discussed superhorizon modes.
Second, we assumed that statistical anisotropy is scale-invariant for simplicity.
However, anisotropic inflation normally predicts scale-dependent anisotropy.
There exist anisotropic attractor solutions, where initially unstable isotropic inflation converges to anisotropic inflation~\cite{Soda:2012zm}.
In this model, the early stage of isotropic inflation is consistent with the observed CMB anisotropies, but sizable anisotropies may exist on small scales.
$g_{\star}(k)$ freezes out when the $k$ mode exits the horizon.
Therefore, the induced tensor powerspectrum would constrain the scale dependence of $g_\star$.
Third, we only considered statistical anisotropies originated from a vector field.
However, remnants of some new spinning particles during inflation possibly produce anisotropic non-Gaussianity~\cite{Arkani-Hamed:2015bza, Franciolini:2017ktv}. 
Extending the present calculation to the higher spins would be straightforward.
In general, the presence of spin~$s$ fields leads to $\ell \leq 2s$ Legendre polynomials in the scalar bispectrum spectrum~\cite{Arkani-Hamed:2015bza, Franciolini:2017ktv}, so we conjecture the induced tensor powerspectrum has $\ell \leq 4s$ statistical anisotropy.
Depending on reheating scenarios, we would be able to constrain the presence of higher-spin fields during inflation severely.
It would also be interesting to consider statistically \textit{isotropic} non-Gaussianity in solid inflation because its squeezed limit formula has quadrupole angular dependence~\cite{Endlich:2012pz, Pajer:2019jhb}.
Forth, we expect similar mode coupling effects for the various gravitational wave events in the early Universe, such as phase transitions, preheating, or topological defects.
Our framework would allow CMB measurements to constrain parameters related to the above events with primordial non-Gaussianity, while we only focused on reheating scenarios and primordial blackhole formations in this paper.

We also leave more general comments on second-order effects of cosmological perturbations.  
One might wonder that similar machinery can be useful to discuss the secondary magnetic fields at a large scale.
However, OPE coefficients in Eq.~\eqref{OPE} cannot be an odd function of $\mathbf k$ due to the permutation symmetry of the operator product.
Hence we cannot include $Y_{1,\pm1}(\hat k)$ in the OPE coefficients, which are essential to get the secondary superhorizon vector perturbations.
Thus, we only have diagrams similar to ($b$) of Fig.~\ref{loop} for the second-order vector perturbations, which have been already discussed in many references~\cite{Matarrese:2004kq,Takahashi:2005nd,Matarrese:2004kq,Kobayashi:2007wd,Fenu:2010kh,Nalson:2013jya,Saga:2015bna,Fidler:2015kkt,Carrilho:2019qlb}.


\begin{acknowledgments}
The author is supported by JSPS Overseas Research Fellowships. 
The author would like to thank Marc Kamionkowski, Cyril Pitrou, Toshiya Namikawa, David Stefanyszyn, Enrico Pajer, Evangelos Sfakianakis, Sebastien Clesse, and Keisuke Inomata for useful discussions and comments.
\end{acknowledgments}

\bibliography{bib}{}
\bibliographystyle{unsrturl}

\end{document}